\documentclass{moriond}

\usepackage{amsmath}
\usepackage{amssymb}

\def\Journal#1#2#3#4{{#1} {\bf #2}, #3 (#4)}
\def\reff@jnl#1{{\rm#1\/}}

\def\mnras{MNRAS}            
\def\ptep{PTEP}            
\def\jcap{JCAP}            
\def\aap{\reff@jnl{A\&A}}               

\def\be{\begin{equation}}
\def\ee{\end{equation}}
\def\bea{\begin{eqnarray}}
\def\eea{\end{eqnarray}}

\newcommand{\Photo}{}

\begin{document}
\vspace*{4cm}
\title{Optimizing blind reconstruction of CMB B-modes for future experiments}

\author{A. Carones}

\address{Dipartimento di Fisica, Università di Roma Tor~Vergata \& INFN Sez. Roma 2, \\
via della Ricerca Scientifica 1, I-00133, Roma, Italy}

\maketitle\abstracts{
The detection of primordial polarization $B$ modes of the Cosmic Microwave Background (CMB) requires exquisite control of Galactic foreground contamination. The Needlet Internal Linear Combination (NILC) method has proven effective in reconstructing CMB $B$ modes without suffering from mis-modeling errors of Galactic emission. 
However, with the most complex foreground models, residual Galactic contamination from NILC is proved to bias, especially at large angular scales, the recovered CMB $B$ modes from simulated data of future CMB experiments. We therefore present two new extensions of NILC, Multi-Clustering NILC (MC-NILC) and optimized constrained Moment ILC (ocMILC), which allow to enhance foreground subtraction in the reconstructed CMB signal.}

\section{Blind reconstruction of CMB $B$ modes}
The detection of primordial polarization $B$ modes of the Cosmic Microwave Background (CMB), targeted by upcoming CMB experiments, is hampered by the contamination of Galactic foregrounds. This points to the need for robust component separation techniques to be applied to future microwave observations. The reference model-independent (\textit{i.e.} blind) cleaning technique is the Needlet Internal Linear Combination \cite{NILC} (NILC). In the NILC pipeline, input multi-frequency $B$-mode maps $B^{i}$ are first decomposed into needlet coefficients $\beta_{j}^{i}$ through a deconvolution for a needlet kernel $\psi_j$: $\beta_{j}^{i}(\hat{\gamma})=\psi_j\circledast B^{i}(\hat{\gamma})$. At each needlet scale $j$, sampling a specific range of angular scales, needlet maps are then linearly combined with frequency-dependent weights $w_{j}^{i}$:
\begin{equation}
\beta_{j}^{NILC}(\hat{\gamma}) = \sum_{i=1}^{N_{\nu}}w_{j}^{i}(\hat{\gamma})\cdot \beta_{j}^{i}(\hat{\gamma}),
\label{eq:NILC}
\end{equation}
so as to recover a blackbody solution $\beta_{j}^{NILC}$ with minimum variance. The derivation of NILC weights,
$\boldsymbol{w_j}(\hat{\gamma}) = \left[\boldsymbol{A}_{\text{CMB}}^{T}C_{j}^{-1}(\hat{\gamma})\,\boldsymbol{A}_{\text{CMB}}\right]^{-1}\boldsymbol{A}_{\text{CMB}}^{T}C_{j}^{-1}(\hat{\gamma})$, requires only knowledge of the CMB spectral energy distribution (SED), $\boldsymbol{A}_{\text{CMB}}$, and the computation of the input needlet covariance matrix: $C_{j}^{ik}(\hat{\gamma})=\langle \beta_{j}^{i}(\hat{\gamma})\cdot \beta_{j}^{k}(\hat{\gamma}) \rangle$, with $\langle \rangle$ a local sample average within Gaussian domains, whose width varies with the needlet scale $j$. The final $B$-mode CMB map, $\tilde{B}_{\text{CMB}}$, is reconstructed through an inverse needlet transform of NILC solutions at the different needlet scales:
\begin{equation}
\tilde{B}_{\text{CMB}}(\hat{\gamma})=\sum_{j}\psi_j\circledast \beta_{j}^{NILC}(\hat{\gamma})=B_{\text{CMB}}(\hat{\gamma}) + B_{\text{fres}}(\hat{\gamma}) + B_{\text{nres}}(\hat{\gamma}),
\label{eq:NILC_out}
\end{equation}
and includes residual contamination by foregrounds ($B_{\text{fres}}$) and instrumental noise ($B_{\text{nres}}$). The contribution of $B_{\text{nres}}$ to the output angular power spectrum can be removed, whereas that of $B_{\text{fres}}$, if relevant at the reionization peak ($\ell \lesssim 10$) or at the recombination bump ($\ell \sim 80$), may bias the estimate of the tensor-to-scalar ratio.

\section{Optimizing blind CMB component separation}
\begin{figure}
\includegraphics[width=0.48\linewidth]{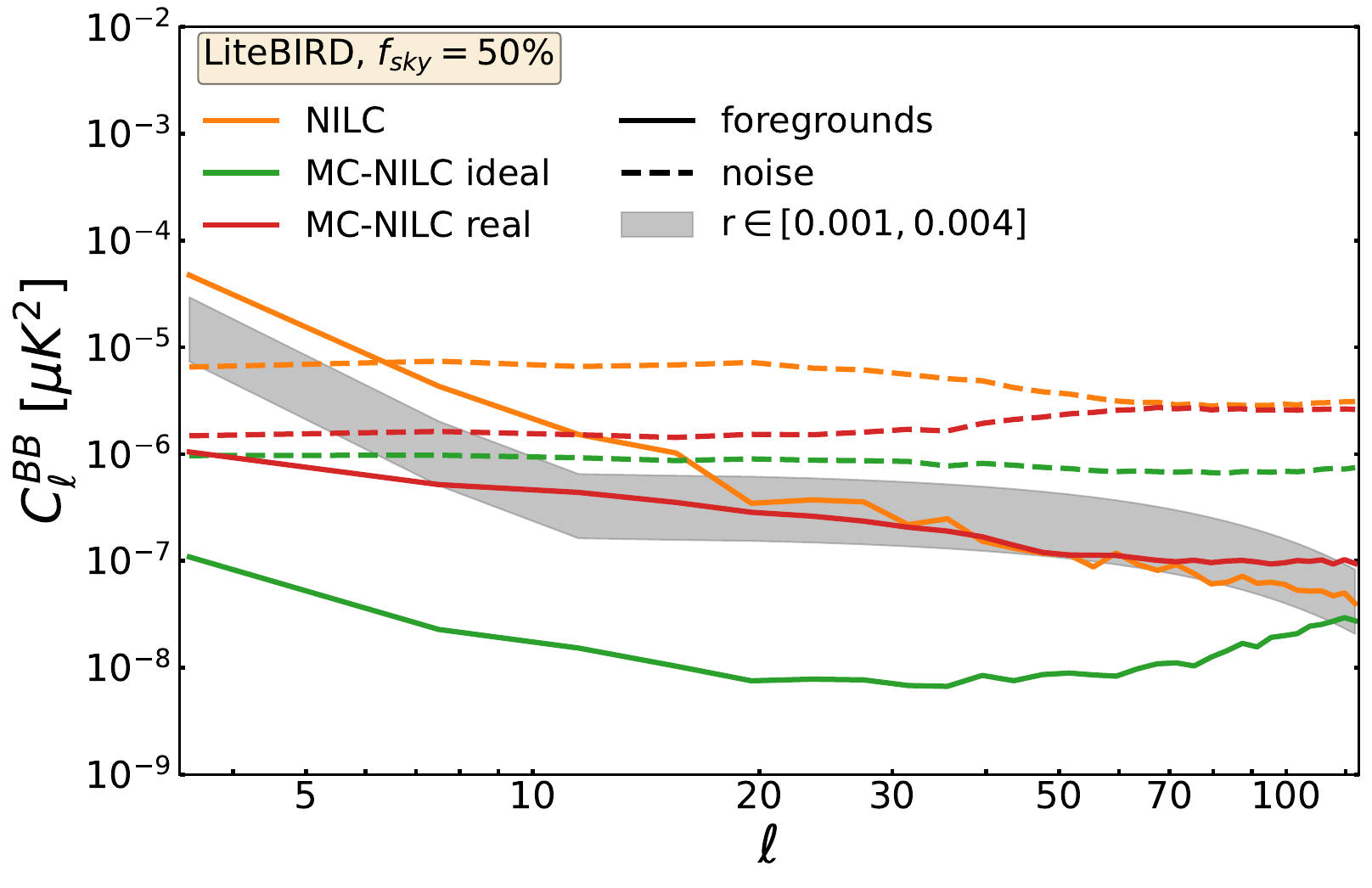}
\includegraphics[width=0.48\linewidth]{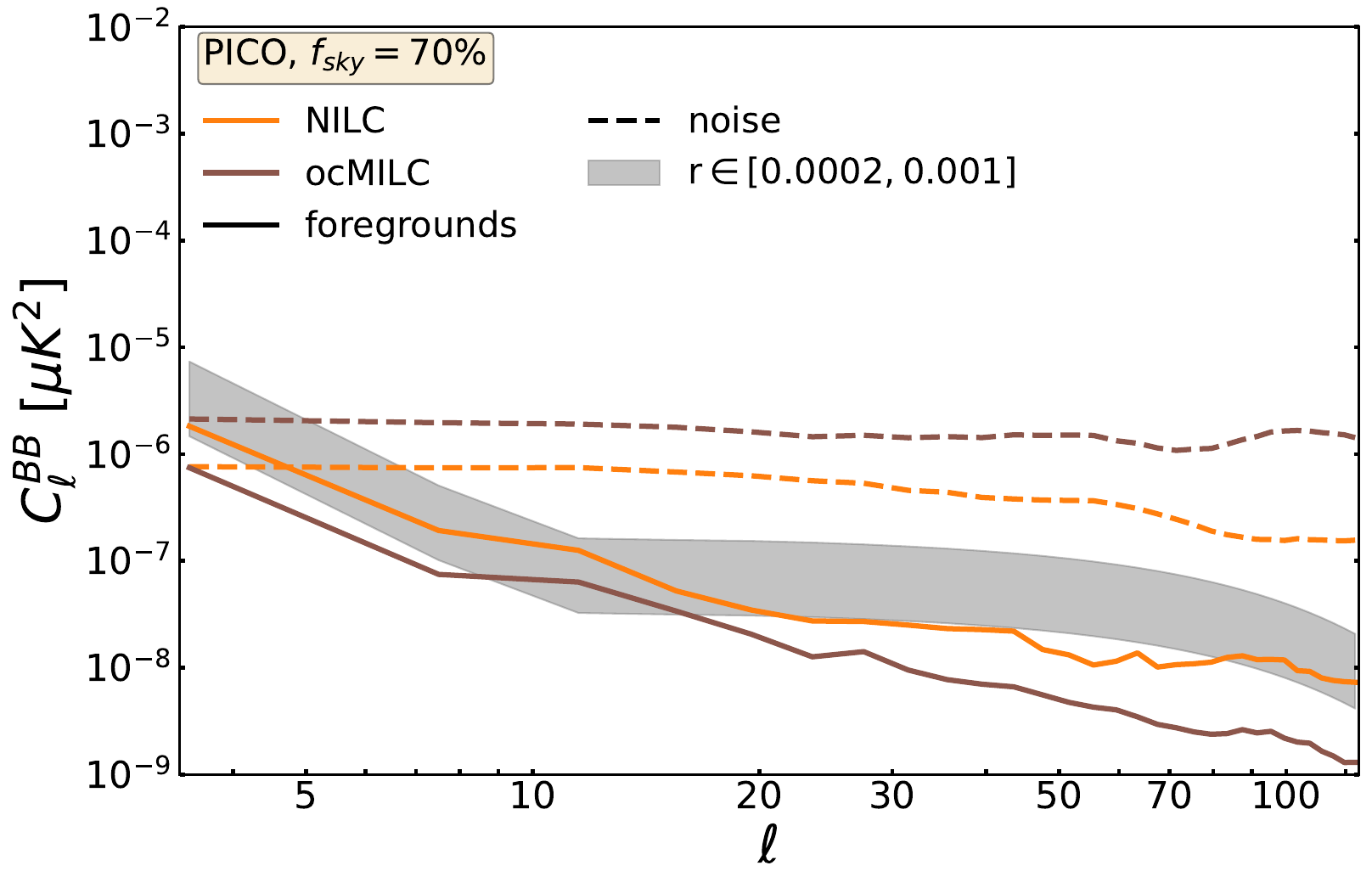}
\caption[]{Angular power spectrum (averaged over $50$ simulations) of foreground (solid) and noise (dashed) residuals in reconstructed CMB $B$-mode maps compared against a reference range of tensor primordial CMB spectra (gray shaded area). Left: results from the application of NILC (orange), MC-NILC with ideal clusters (green), and MC-NILC with real clusters (red) on \textit{LiteBIRD} sky simulations, with spectra computed in a sky fraction $f_{sky}=50\%$. Left: results from the application of NILC (orange) and ocMILC (brown) on \textit{PICO} sky simulations with spectra computed in a sky fraction $f_{sky}=70\%$. Power spectra are binned with $\Delta \ell = 4$.}
\label{fig:cls}
\end{figure}

We consider simulations of \textit{LiteBIRD} \cite{PTEP} and \textit{PICO} \cite{PICO} satellite experiments, targeting a sensitivity on the tensor-to-scalar ratio of $\sigma(r)=0.001$ and $0.0002$, respectively. Galactic foreground emission is simulated with the \texttt{PySM} python package~\cite{pysm} assuming the dust \texttt{d1} model (modified blackbody SED) and synchrotron \texttt{s1} model (power-law SED), both featuring spatial variations of the corresponding spectral parameters. After proper masking, at large angular scales, the NILC foreground residuals still have power comparable to that of the targeted amplitude of tensor modes, as shown by solid orange lines in the left (\textit{LiteBIRD}) and right (\textit{PICO}) panels of Fig.~\ref{fig:cls}. As demonstrated in \cite{MCNILC} and \cite{ocMILC}, these residuals would produce a non-negligible bias on the estimated tensor-to-scalar ratio. We thus present two alternative optimizations of NILC: Multi-Clustering NILC \cite{MCNILC} (MC-NILC) and optimized constrained moment ILC \cite{ocMILC} (ocMILC). 

The MC-NILC technique implements variance minimization in an optimized sky partition with each patch including only pixels with similar spectral properties of the input $B$-mode foregrounds. This optimization allows variance minimization to handle a simpler Galactic emission in each region. 
Foreground spectral properties are blindly estimated by computing the ratio $\tilde{B}^{\text{hf}}_{\text{fgds}}/\tilde{B}^{\text{cf}}_{\text{fgds}}$ of templates of $B$-mode foreground emission, one at a high-frequency channel, while the other at a central "CMB" frequency \cite{MCNILC}. We report results for an ideal case, where the templates correspond to the input foreground maps, and for a realistic one, with templates derived from input noisy data by applying the Generalized Needlet ILC technique \cite{GNILC} (GNILC). MC-NILC pipeline has been validated on \textit{LiteBIRD} sky simulations and is proved to significantly reduce Galactic contamination, especially at large angular scales, as shown in the left panel of Fig.~\ref{fig:cls}. 

The ocMILC technique, instead, enhances foreground subtraction by deprojecting some moments of the polarized foreground emission \cite{moms} through the linear combination in Eq.~\ref{eq:NILC}. This deprojection, already introduced in \cite{cMILC}, is optimized in ocMILC. First, the optimal number of moments to deproject across the sky and needlet scales is derived through a blind diagnostic of input foreground complexity. The optimal set of moments and the amount of deprojection of each moment is then derived by considering the combination leading to the minimum denoised output variance $\boldsymbol{w^j}^{T} \left(C_j\, - N_j\right)\, \boldsymbol{w^j}$, with $N_j$ the input noise needlet covariance \cite{ocMILC}. The effectiveness of the ocMILC pipeline in reducing Galactic contamination at all angular scales, with only a moderate increase of noise residuals, is assessed on \textit{PICO} sky simulations and shown in the right panel of Fig.~\ref{fig:cls}.

Both the MC-NILC and ocMILC methods are robust to a variety of different foreground models and proved to provide unbiased estimates of the tensor-to-scalar ratio with associated uncertainties within the targeted sensitivity of the considered experiments \cite{MCNILC,ocMILC}.


\section*{References}


\begin{thebibliography}{99}

\bibitem{NILC} J.Delabrouille et al., \Journal{\aap}{493}{835-857}{2009}. 
    
\bibitem{PTEP} LiteBIRD Collaboration, \Journal{\ptep}{4}{042F01}{2023}. 

\bibitem{PICO} R.Aurlien et al., \Journal{\jcap}{2023}{034}{2023}. 

\bibitem{pysm} A.Zonca et al., The Journal of Open Source Software \textbf{6}, 3783 (2021). 



\bibitem{MCNILC}A.Carones et al., \Journal{\mnras}{525}{3117-3135}{2023}. 

\bibitem{ocMILC} A.Carones and M.Remazeilles, \textit{arXiv e-prints}: 2402.17579[astro-ph.CO]. 

\bibitem{GNILC} M.Remazeilles, et al., \Journal{\mnras}{418}{467}{2011}. 

\bibitem{moms} J.Chluba et al., \Journal{\mnras}{472}{1195}{2017}. 

\bibitem{cMILC} M.Remazeilles et al., \Journal{\mnras}{503}{2478}{2021}. 


\end{thebibliography}
\end{document}